\documentclass[11pt,a4paper]{article}

\usepackage{amsmath,amssymb,amsthm}
\usepackage{graphicx}
\usepackage{physics}
\usepackage{enumitem}
\usepackage{xspace}

\usepackage{orcidlink}

\usepackage[margin=1in]{geometry}
\usepackage{url}
\usepackage{hyperref}
\usepackage{color}
\definecolor{refcolor}{RGB}{160,35,0}
\definecolor{hrefcolor}{RGB}{0,35,190}
\hypersetup{
    colorlinks,
    citecolor=refcolor,
    filecolor=refcolor,
    linkcolor=hrefcolor,
    urlcolor=hrefcolor
}

\def\({\left(}
\def\){\right)}
\def\[{\left[}
\def\]{\right]}

\newcommand{\hilbert}{\mathcal{H}}
\newcommand{\pobs}[1]{#1}
\newcommand{\obs}[1]{\mathsf{\pobs{#1}}}

\newcommand{\R}{\mathbb{R}}

\newcommand{\Z}{\mathbb{Z}}

\newcommand{\WdWSpaceM}{{\mkern-6mu}}
\newcommand{\WdWSpaceMM}{{\mkern-9mu}}
\newcommand{\WdWSpaceR}{{\mkern-8mu}}

\newcommand{\kett}[1]{\ket{\WdWSpaceM\ket{#1}\WdWSpaceR}}

\newcommand{\brakett}[2]{\left.\bra{#1}{\WdWSpaceMM}\ket{#2}{\WdWSpaceR}\right\rangle}

\newcommand{\q}{\boldsymbol{q}}

\newcommand{\M}{\mathbf{M}}

\newtheorem{assumption}{Assumption}
\newtheorem{insight}{Insight}
\newtheorem{principle}{Principle}
\newtheorem{problem}{Problem}

\theoremstyle{remark}

\newtheoremstyle{nospace} % name of the style
  {3pt}                   % Space above
  {3pt}                   % Space below
  {\itshape}              % Body font
  {}                      % Indent amount
  {\bfseries}             % Theorem head font
  {.}                     % Punctuation after theorem head
  {.5em}                  % Space after theorem head
  {\thmname{#1}\thmnumber{#2}\thmnote{ (#3)}} % <--- HEADER SPECIFICATION

\theoremstyle{nospace}

\newtheorem{defCustom}{}
\makeatletter
\newcommand{\setdefCustomtag}[1]{% \setdefCustomtag{<tag>}
  \let\oldthedefCustom\thedefCustom% Store \thedefCustom
  \renewcommand{\thedefCustom}{{\normalfont\textbf{#1}}}% Redefine it to a fixed value
  \g@addto@macro\enddefCustom{% At \end{defCustom}, ...
    \global\let\thedefCustom\oldthedefCustom}% ...restore \thetheorem
  }
\makeatother

\allowdisplaybreaks

\begin{document}

\title{The preferred-time problem in the conditional interpretation of time}
\author{\orcidlink{0000-0002-2765-1562}\ O.C. Stoica~\footnote{Dept. Th. Physics, NIPNE-HH, Bucharest, Romania. \href{mailto:cristi.stoica@theory.nipne.ro}{cristi.stoica@theory.nipne.ro},  \href{mailto:holotronix@gmail.com}{holotronix@gmail.com}}}

%todo
\maketitle

\begin{abstract}
In timeless formulations of quantum theory, the Page-Wootters proposal (PW) recovers time and dynamics as relative clock-and-world states. By interpreting the total timeless system as a clock entangled with the world, the states having a definite clock reading appear to recover the temporal states of the world as relative states, and dynamics emerges. But an entangled state admits infinitely many decompositions as a superposition of product states, each decomposition corresponding to a different but equally valid time operator of the same clock system. Different valid choices of the time operator for the same clock system lead to different physical histories, where the world states in a history are superpositions of the world states from another history. Fixing one operator as ``the'' clock time would import the temporal meaning that the Page-Wootters proposal is meant to explain in the first place. Therefore, the interpretation of PW that the world state is passively conditioned on the clock state cannot hold. The timeless state is resolved into time-dependent states by the intrinsic pointer observables of the world, without having to outsource the role of the time operator to a separate clock. Fortunately, the formalism of PW remains valid, provided that it is interpreted in terms of intrinsic pointer observables of the world itself.
\end{abstract}

%todo \maketitle


\begin{quote}
\emph{Time is what keeps everything from happening at once.}
\end{quote}
\begin{flushright}
Ray Cummings, \emph{The Girl in the Golden Atom}, 1919.
\end{flushright}

%------------------------------------------------------------%
\section{Introduction}
\label{s:intro}

Page and Wootters proposed a formalism that can describe a quantum world in a timeless fashion~\cite{PageWootters1983EvolutionWithoutEvolution,Page1986DensityMatrixOfTheUniverse,Page1994ClockTimeAndEntropy,Wootters1984TimeReplacedByQuantumCorrelations}.
A way to extract time evolution from a static system seems to be needed because gravitational coupling induces a superselection rule for energy, analogous to the charge superselection rule associated with long-range electromagnetic interactions~\cite{WickWightmanWigner1952TheIntrinsicParityOfElementaryParticles}.
This is explicitly visible in canonical quantum gravity, where the solutions to the Wheeler-DeWitt constraint equation are stationary~\cite{Dewitt1967QuantumTheoryOfGravityI_TheCanonicalTheory,Dewitt1967QuantumTheoryOfGravityII_TheManifestlyCovariantTheory},
\begin{equation}
\label{eq:PW-WDW}
\obs{H}\kett{\Psi}=0.
\end{equation}

In fact, the Wheeler-DeWitt constraint equation~\eqref{eq:PW-WDW} appears even if we do not assume gravity~\cite{Stoica2026QuantumTimeWithoutNegativeEnergy}, simply by taking seriously the existence of change even if energy positivity seems to forbid it due to a Lemma by Hegerfeldt and Ruijsenaars~\cite{HegerfeldtRuijsenaars1980RemarksOnCausalityLocalizationAndSpreadingOfWavePackets}.

The Page-Wootters proposal (PW) aims to recover time and dynamics by relying on the existence of a clock subsystem $c$ and a ``rest of the world'' subsystem $r$, with total Hilbert space $\hilbert=\hilbert_c\otimes\hilbert_r$.
The total Hamiltonian is
\begin{equation}
\label{eq:PW-noninteracting}
\obs{H}=\obs{H}_c\otimes\obs{I}_r+\obs{I}_c\otimes\obs{H}_r,
\end{equation}
where the clock Hamiltonian $\obs{H}_c$ is
\begin{equation}
\label{eq:clock-H}
\obs{H}_c=-i\hbar\pdv{\tau}.
\end{equation}

According to PW, the states satisfying the constraint equation~\eqref{eq:PW-WDW} are of the form
\begin{equation}
\label{eq:PW-state-timeless}
\kett{\Psi}=\int_\R\ket{\tau}_c\ket{\psi(\tau)}_r\,\dd \tau,
\end{equation}
where $\ket{\tau}_c$ are generalized eigenstates of a ``time operator'' $\obs{T}_c=\int_{\R}\tau\dyad{\tau}_c\,\dd \tau$ canonically conjugate to $\obs{H}_c$,
\begin{equation}
\label{eq:clock-time-operator}
[\obs{T}_c,\obs{H}_c]=i\hbar.
\end{equation}

The time operator $\obs{T}_c$ does not have an \emph{a priori} meaning of time, this meaning seeming to be acquired from the embedding of a history $\ket{\psi(\tau)}_r$ in equation~\eqref{eq:PW-state-timeless}.

While the timeless state $\kett{\Psi}$ is not normalizable, Page and Wootters interpret the world state $\ket{\psi(\tau)}_r$ as conditioned on the clock state being $\ket{\tau}_c$.
The usual expectation value of an observable $\obs{A}_r$ is obtained as a \emph{conditional expectation value} of the stationary observable $\obs{A}=\obs{I}_c\otimes\obs{A}_r$ given that the clock reads $\tau$ as
\begin{equation}
\label{eq:conditional-expectation-value}
E(\obs{A}|\tau)=\frac{\tr(\obs{A}\obs{P}_\tau\rho)}{\tr(\obs{P}_\tau\rho)},
\end{equation}
where $\obs{P}_\tau:=\dyad{\tau}\otimes\obs{I}_r$.
The Schr\"odinger equation for $\ket{\psi(\tau)}_r$ can also be recovered from equation~\eqref{eq:PW-WDW}~\cite{Wootters1984TimeReplacedByQuantumCorrelations}.

The Page-Wootters proposal was analyzed and developed in various directions. Starting from Kucha\v{r}'s criticism~\cite{Kuchar2011TimeAndInterpretationsOfQuantumGravity}, several reconstructions and responses were proposed~\cite{GiovannettiLloydMaccone2015QuantumTime,LoveridgeMiyadera2019RelativeQuantumTime,MacconeSacha2020QuantumMeasurementsOfTime,CastroruizGiacominiBelenchiaBrukner2020QuantumClocksAndTemporalLocalisabilityOfEventsInThePresenceOfGravitatingQuantumSystems,FotiEtAl2021TimeAndClassicalEqOfMotionFromQEntanglementViaPageWoottersGeneralizedCoherentStates,HohnSmithLock2021TrinityOfRelationalQuantumDynamics,Gambini2022SolutionProblemOfTimeQuantumGravityAlsoTimeArrivalProblemQM,Rijavec2022HeisenbergPictureEvolutionWithoutEvolution,Rijavec2023robustnessOfThePageWoottersConstructionAcrossDifferentPicturesStatesOfTheUniverseAndSystemClockInteractions,SuleymanovCohen2023QuantumFramesOfReferenceAndRelationalFlowOfTime,RidleyCohen2025TwoTimesOrNone}, and even experimental models~\cite{MorevaBridaGramegnaGiovannettiMacconeGenovese2014TimeFromQuantumEntanglementAnExperimentalIllustration,MorevaGramegnaBridaMacconeGenovese2017QuantumTimeExperimentalMultitimeCorrelations}. Other good analyses or reviews can be found in~\cite{Adlam2022WatchingTheClocksInterpretingPageWoottersFormalismAndInternalQRF,AltaieBeigeHodgson2022TimeAndQuantumClocksAReviewOfRecentDevelopments}.

In the following I will analyze two key assumptions on which the PW proposal is normally understood:
\begin{assumption}
\label{assumption:timelessness}
The total system $c+r$ is static, it does not presuppose time or dynamics. Time and dynamics emerge solely from the entanglement between the clock and the rest of the world.
\end{assumption}

\begin{assumption}
\label{assumption:clock-conditions}
The state of the world at time $\tau$ is conditioned on the clock's state $\ket{\tau}_c$,
\begin{equation}
\label{eq:relative-state-c-r}
\ket{\tau}_c\mapsto\ket{\psi(\tau)}_r.
\end{equation}
\end{assumption}

I show a tension between these two assumptions. Assumption~\ref{assumption:timelessness} says that no temporal meaning is available in advance, while Assumption~\ref{assumption:clock-conditions} treats one particular clock basis as already the basis of clock readings, at the same time confering an active role of the clock state $\ket{\tau}_c$ to determine the world state $\ket{\psi(\tau)}_r$. The question is what, inside the timeless theory, selects that specific basis rather than another equally valid basis. There are already difficulties in selecting the clock subsystem from the total clock-world system, due to the \emph{clock ambiguity problem}~\cite{Albrecht1995TheoryOfEeverythingVsTheoryOfAnything,AlbrechtIglesias2008ClockAmbiguityAndTheEmergenceOfPhysicalLaws}, whose range was extended in~\cite{Stoica2026TheClockAmbiguityProblemExtendedOrExtinguished}. But the problem investigated here occurs in addition to fixing the clock subsystem.

It is useful to understand a key insight on which Page and Wootters based the argument that their proposal is a valid way to recover the quantum dynamics and time. They consider~\cite{Wootters1984TimeReplacedByQuantumCorrelations,Page1994ClockTimeAndEntropy} their proposal to be an application of Everett's idea of \emph{relative states}~\cite{Everett1957RelativeStateFormulationOfQuantumMechanics,Everett1973TheTheoryOfTheUniversalWaveFunction}.
Page's reply to Kucha\v{r}'s criticism was explicit about this~\cite{Page1994ClockTimeAndEntropy}:

\begin{quote}
\textbf{Karel Kucha\v{r}:} \emph{Don, you are the first person I met who simultaneously believes in the existence of many worlds and is a solipsist of an instant.}

\textbf{Don Page:} \emph{I believe that different instants, i.e., different clock times, are actually examples of the different worlds. They all exist, but each observation, and its associated conditional probability, occurs at one single time (assuming that the condition includes or implies a precise value of the clock time in question). We can only directly observe and
be aware of the world, and the time, in and at which we exist, though the correlations in memories and other structures we observe in one world give indirect evidence of other worlds, and other clock times, in the full quantum state of the universe. [...]}
\end{quote}

So the basic insight at the foundation of the PW proposal can be summarized as
\begin{insight}
\label{insight:relative-state}
The idea of relative states, used by Everett to address the measurement problem, can also explain how different times are obtained from a timeless state: the timeless state decomposes into states at different times similar to how the universal wavefunction $\psi(t)$ decomposes into branches in Everett's proposal~\cite{Everett1957RelativeStateFormulationOfQuantumMechanics,Everett1973TheTheoryOfTheUniversalWaveFunction}.
\end{insight}

In Section~\ref{s:preferred-time-problem} we will see that Assumption~\ref{assumption:timelessness} implies that there are infinitely many equally valid ways to choose the time observable $\obs{T}_c$ for the same clock.
Since time is not assumed to begin with, but it has to emerge, there is no way to pick a preferred time observable $\obs{T}_c$.

This, in turn, implies that there are infinitely many equally valid but physically distinct ways to decompose $\kett{\Psi}$ into a superposition of product states of the form~\eqref{eq:PW-state-timeless}, all compatible with the same tensor-product decomposition $\hilbert=\hilbert_c\otimes\hilbert_r$ and with the PW proposal.
This observation and Assumption~\ref{assumption:clock-conditions} imply that the instantaneous states of the world are ambiguously defined.
Assumption~\ref{assumption:timelessness} implies that these different time observables are all equally valid, because whatever time is, it cannot be assumed in advance to be special compared with other observables.
Note that this problem does not arise if we assume that time is fundamental and the clock's state and the world's state really change. But this amounts to rejecting Assumption~\ref{assumption:timelessness}, and the whole rationale of the PW proposal is to recover time without assuming it \emph{a priori}.

I call the ambiguity introduced in Section~\ref{s:preferred-time-problem} \emph{the preferred-time problem of clocks}, to distinguish it from the \emph{clock ambiguity problem}~\cite{Albrecht1995TheoryOfEeverythingVsTheoryOfAnything,AlbrechtIglesias2008ClockAmbiguityAndTheEmergenceOfPhysicalLaws} or its extended version~\cite{Stoica2026TheClockAmbiguityProblemExtendedOrExtinguished}. The clock ambiguity problem, in its original and extended versions, is based on choosing different tensor product decompositions $\hilbert\cong\hilbert_c'\otimes\hilbert_r'$. In~\cite{Stoica2026TheClockAmbiguityProblemExtendedOrExtinguished} it was shown that taking the physical meanings of observables into account resolves the clock ambiguity problem.
However, this cannot solve the new ambiguity introduced in Section~\ref{s:preferred-time-problem}, namely the preferred-time problem of clocks, because if we assume that a particular operator $\obs{T}_c$ already has the physical meaning of a time operator, we smuggle in the very meaning that the PW proposal aims to recover as emergent.

In Section~\ref{s:clock-dont-control-time} we will see that the modern version of Everett's idea that grounds the whole PW proposal (Insight~\ref{insight:relative-state}) resolves the ambiguity, but in doing so, it replaces the conditioning from equation~\eqref{eq:relative-state-c-r} with its reverse conditioning,
\begin{equation}
\label{eq:relative-state-r-c}
\ket{\psi(\tau)}_r\mapsto\ket{\tau}_c.
\end{equation}

In fact, it should already be clear that, contrary to Assumption~\ref{assumption:clock-conditions}, the world's states condition the clock states, and not the other way around, because a system cannot gain its interpretation as a clock merely due to the specific form of entanglement in which partakes, but because it tracks change in a physically meaningful way. This is especially clear for a noninteracting Hamiltonian such as~\eqref{eq:PW-noninteracting}, which contains no causal effect whatsoever from the clock to the rest of the world. But even an interaction that allows the clock to be read does not by itself justify the conclusion that what was read was time.
Whatever time is, it happens within the world, and the clock only measures it, it does not control it.
Assumption~\ref{assumption:clock-conditions} is a theoretician's description, a \emph{God's-eye} perspective of someone who already knows or assumes $\obs{T}_c$ as the time observable, as opposed to the view of an observer immersed in the world. An immersed observer cannot access a noninteracting clock or be actively conditioned by it. Even if the clock is made accessible through an interaction, the observer still has no independent grounds for treating it as the clock that controls the dynamics of processes within the world, if time is purely emergent.
An observer can access nothing more than the macrostate, consisting of conglomerates of relative macroscopic pointer states. These are sufficient to separate not only the branches of the universal wavefunction, but also the different stages of the history, in other words, instants of time.

But then, what job remains for the clock subsystem? It does its job as a bookkeeping device for the theoretician's God's eye, but the real job of defining time is done by the time observable intrinsic to the world. Equation~\eqref{eq:relative-state-r-c} collapses into the tautology
\begin{equation}
\label{eq:relative-state-r}
\ket{\psi(\tau)}_r\mapsto\ket{\psi(\tau)}_r
\end{equation}
as time bootstraps itself from within the world.
The timeless world $\kett{\Psi}$ can therefore be written simply as
\begin{equation}
\label{eq:bootstrap-state-timeless}
\kett{\Psi}=\int_\R\ket{\psi(\tau)}_r\,\dd \tau.
\end{equation}

That it still satisfies the Wheeler-DeWitt constraint equation $\obs{H}\kett{\Psi}=0$ it was shown to follow from the invariance of $\int_\R\ket{\psi(\tau)}_r\,\dd \tau$ under translations in $\tau$, under the right regularization and boundary assumptions~\cite{Stoica2026QuantumTimeWithoutNegativeEnergy}. The clock subsystem is rendered redundant and inconsequential.

Section~\ref{s:PW-reinterpret} argues that we can still use the formalism of the Page-Wootters proposal, provided that we interpret the decomposition $\hilbert=\hilbert_c\otimes\hilbert_r$ as separating degrees of freedom intrinsic to the world, rather than as decomposing the total system into a clock subsystem and the rest of the world.

%------------------------------------------------------------%
\section{The preferred-time problem of clocks}
\label{s:preferred-time-problem}

Briefly, the preferred-time problem is:
\begin{problem}
\label{problem:preferred-time-problem}
If no temporal meaning is presupposed, then the clock does not determine which of its canonically conjugate observables is the physical time observable.
\end{problem}

According to Assumption~\ref{assumption:timelessness}, time and dynamics emerge due to the clock observables $\obs{H}_c$ and $\obs{T}_c$.
Time emerges because the eigenstates of $\obs{T}_c$ separate $\kett{\Psi}$ into a particular superposition of relative states of the form~\eqref{eq:PW-state-timeless}. The clock's dynamics is generated by the Hamiltonian $\obs{H}_c$.
But interpreting $\obs{H}_c$ and $\obs{T}_c$ as Hamiltonian and clock observable assumes both dynamics and time.
One may hope that $\obs{H}_c$ and $\obs{T}_c$ are the unique observables satisfying the canonical commutation relation $[\obs{T}_c,\obs{H}_c]=i\hbar$, and it may even be tempting to claim that the Stone-von Neumann theorem proves exactly this uniqueness~\cite{Stone1930LinearTransformationsInHilbertSpaceIIIOperationalMethodsAndGroupTheory,Stone1932OnOneParameterUnitaryGroupsInHilbertSpace,vonNeumann1931DieEindeutigkeitDerSchrodingerschenOperatoren,vonNeumann1932UberEinenSatzVonHerrnMHStone,Hall2013QuantumTheoryForMathematicians}. The Stone-von Neumann theorem gives uniqueness up to unitary equivalence, but it does not say that every unitarily equivalent pair gives the same physical description of the world, which would be false~\cite{Stoica2022SpaceThePreferredBasisCannotUniquelyEmergeFromTheQuantumStructure,Stoica2026NoChangeInHilbertSpaceFundamentalism,Stoica2026CommentOnOnTheEmergenceOfPreferredStructuresInQuantumTheoryBySoulasFranzmannAndDiBiagio,Stoica2026TheClockAmbiguityProblemExtendedOrExtinguished}.

Indeed, any unitary transformation $\obs{S}_c$ of $\hilbert_c$ can give different operators $\obs{H}_c'=\obs{S}_c\obs{H}_c\obs{S}_c^{-1}$ and  $\obs{T}_c'=\obs{S}_c\obs{T}_c\obs{S}_c^{-1}$ satisfying the same commutation relation $[\obs{T}_c',\obs{H}_c']=i\hbar$.
Then, the timeless state can be expressed in infinitely many ways as a clock and the rest of the world.
None of these choices within $\hilbert_c$ is a poorer clock, contrary to a claim from~\cite{MarlettoVedral2017EvolutionWithoutEvolutionAndWithoutAmbiguities}, nor are they physically equivalent merely because they are unitarily related.

The symmetry transformation given by any unitary $\obs{S}_c$ gives a new generalized clock basis, which I denote by $\ket{\vartheta}_c$. Equivalently, when the spectra are identified, one may write $\ket{\vartheta}_c=\obs{S}_c\ket{\tau=\vartheta}_c$. Substituting
\begin{equation}
\label{eq:clock-change-basis}
\ket{\tau}_c=\int_\R\braket{\vartheta}{\tau}_c\ket{\vartheta}_c\,\dd\vartheta
\end{equation}
in equation~\eqref{eq:PW-state-timeless}, we get
\begin{equation}
\label{eq:PW-state-timeless-time-map-new-calc}
\begin{array}{rcl}
\kett{\Psi}
&\stackrel{\eqref{eq:PW-state-timeless}}{=}&\int_\R\ket{\tau}_c\ket{\psi(\tau)}_r\,\dd \tau\\
&\stackrel{\eqref{eq:clock-change-basis}}{=}&\int_\R\(\int_\R\braket{\vartheta}{\tau}_c\ket{\vartheta}_c\,\dd\vartheta\)\otimes\ket{\psi(\tau)}_r\,\dd\tau \\
&\stackrel{\phantom{\eqref{eq:clock-change-basis}}}{=}&\int_\R\ket{\vartheta}_c\otimes\(\int_\R\braket{\vartheta}{\tau}_c\ket{\psi(\tau)}_r\,\dd\tau\)\,\dd\vartheta\\
&\stackrel{\eqref{eq:PW-world-state-map-new}}{=}&\int_\R\ket{\vartheta}_c\ket{\psi'(\vartheta)}_r\,\dd\vartheta
\end{array}
\end{equation}
where
\begin{equation}
\label{eq:PW-world-state-map-new}
\ket{\psi'(\vartheta)}_r
=\int_\R\braket{\vartheta}{\tau}_c\ket{\psi(\tau)}_r\,\dd\tau.
\end{equation}

Therefore, each $\vartheta$-state of the rest of the world is described, relative to $\ket{\vartheta}_c$, as a superposition of the original states $\ket{\psi(\tau)}_r$ relative to $\ket{\tau}_c$.

Thus the state of the world and its history can become completely different if we choose a different time operator in the clock system.

One may hope that requiring that the Wheeler-DeWitt equation~\eqref{eq:PW-WDW} remains true fixes the operators $\obs{H}_c$ and $\obs{T}_c$.
But even if we impose the stronger condition that the transformation $\obs{S}_c$ leaves the Hamiltonian $\obs{H}_c$ unchanged, there are still infinitely many choices for $\obs{S}_c$ that commute with $\obs{H}_c$ but change the basis $\(\ket{\tau}_c\)_{\tau}$ nontrivially.
All operators of the form $\obs{T}_c'=\obs{T}_c+f\(\obs{H}_c\)$ satisfy $[\obs{T}_c',\obs{H}_c]=i\hbar$.
Given that real analytic functions $f$ form an infinite-dimensional space, there are infinitely many distinct ways to choose the time operator.
Among these, only the constant ones, $f\(\obs{H}_c\)=-a\obs{I}_c$, give a time operator $\obs{T}_c'$ with the same eigenvectors as $\obs{T}_c$, with eigenvalues differing only by a time translation with $a$.
All other choices give nonequivalent solutions having the generic form~\eqref{eq:PW-world-state-map-new}.

Let us call this problem \emph{the preferred-time problem of clocks}.
This is not merely a mathematical nonuniqueness problem. The mathematical nonuniqueness would be harmless if all choices corresponded to the same macroscopic history. But two generic admissible time observables of the same clock give physically different sequences of instantaneous worlds.
A superposition of a state of the world as it is today and a state of the world as it was yesterday is physically different from the today's state.

%------------------------------------------------------------%
\section{Clocks do not control time, they measure it}
\label{s:clock-dont-control-time}

To understand the solution of the preferred-time problem of clocks, it is useful to revisit the modern accounts of Everett's insight of relative states~\cite{Everett1957RelativeStateFormulationOfQuantumMechanics,Everett1973TheTheoryOfTheUniversalWaveFunction}. 
This is in the spirit of the Page-Wootters proposal, as seen in Section~\ref{s:intro}, but updated with the understanding gained in the meantime.
This is not a commitment to Everett's interpretation of quantum mechanics, since this insight of Everett's is common to all interpretations. This universality can be illustrated by the measurement problem.
According to the Schr\"odinger equation, the measurement of the spin of a spin-$1/2$ particle along the $z$-axis results in the superposition
\begin{equation}
\label{eq:measurement-problem}
a\ket{\uparrow}_s\ket{\text{``$\uparrow$''}}_p+b\ket{\downarrow}_s\ket{\text{``$\downarrow$''}}_p,
\end{equation}
where $\ket{\uparrow}_s$ and $\ket{\downarrow}_s$ are states of the observed system, $\ket{\text{``$\uparrow$''}}_p$ and $\ket{\text{``$\downarrow$''}}_p$ are states of the pointer indicating the outcome of the measurement, and $\abs{a}^2+\abs{b}^2=1$.
The Schr\"odinger unitary evolution predicts that, after the measurement, the observed system and the measuring apparatus are in the entangled state from equation~\eqref{eq:measurement-problem}.

On the other hand, we always observe only one of the terms in the superposition~\eqref{eq:measurement-problem}.
If we include an observer as a third subsystem, with possible states $\ket{\text{I see ``$\uparrow$''}}_o$ and $\ket{\text{I see ``$\downarrow$''}}_o$, the entangled state~\eqref{eq:measurement-problem} extends to 
\begin{equation}
\label{eq:measurement-problem-obs}
a\ket{\uparrow}_s\ket{\text{``$\uparrow$''}}_p\ket{\text{I see ``$\uparrow$''}}_o+b\ket{\downarrow}_s\ket{\text{``$\downarrow$''}}_p\ket{\text{I see ``$\downarrow$''}}_o.
\end{equation}

The insight at the origin of Everett's proposal is that observers register only definite pointer states and, due to the linearity of quantum mechanics, each observer state has access only to one branch of the entangled state~\eqref{eq:measurement-problem-obs}.
The idea of relative states allows this to extend to many observers reading the same pointer state, to Alice and Bob doing the EPR experiment~\cite{EPR35,Bohm1951TheParadoxOfEinsteinRosenAndPodolsky} and so on, resulting in the commonly-agreed classical world familiar to us at the macroscopic level.

The above discussion is not committed to Everett's interpretation. Indeed, Bohm's pilot-wave theory already included a similar account, for the wavefunction playing the role of a pilot wave~\cite{Bohm1952SuggestedInterpretationOfQuantumMechanicsInTermsOfHiddenVariables}. He needed this so that the point particles follow one branch rather than another, without later interference between branches that would spoil the Born-rule account.
Both Bohm and Everett described quite eloquently in their works what later became known as decoherence. Collapse theories also rely on this decomposition until the wavefunction is spontaneously localized, picking one alternative rather than another, again for compatibility with the Born rule~\cite{GhirardiRiminiWeber1986GRWInterpretation}.

There is no unique way to write the state as a superposition as in equations~\eqref{eq:measurement-problem} or~\eqref{eq:measurement-problem-obs}.
Assuming the subsystem spaces fixed, the decompositions~\eqref{eq:measurement-problem} and~\eqref{eq:measurement-problem-obs} are determined by the basis $\big(\ket{\text{``$\uparrow$''}}_p,\ket{\text{``$\downarrow$''}}_p\big)$ of the measuring device, and by the observer's basis $\big(\ket{\text{I see ``$\uparrow$''}}_o,\ket{\text{I see ``$\downarrow$''}}_o\big)$.
Choose a different basis, and the same state is expressed as a different superposition.
This ambiguity needs to be resolved, otherwise there would be an inflation of observer states entangled with pointer states and observed-system states.

The first major update of this account of Everett's relative states assumed that preferred bases are selected dynamically from all possible bases~\cite{Zeh1996DecoherenceAndTheAppearanceOfAClassicalWorldInQuantumTheory,zurek1981PointerBasisOfQuantumApparatus}.
While the requirement that the pointer states are stable reduces the set of bases that can play this role, it does not lead to a unique result~\cite{Stoica2022SpaceThePreferredBasisCannotUniquelyEmergeFromTheQuantumStructure,Stoica2026CommentOnOnTheEmergenceOfPreferredStructuresInQuantumTheoryBySoulasFranzmannAndDiBiagio}. Regardless of the structural or dynamical conditions imposed, too many bases satisfying those conditions remain, so that the correlation between the pointer states and the observer's own records of the pointer states is exactly $0$~\cite{Stoica2023AreObserversReducibleToStructures}. For timeless systems as in the PW proposal this ambiguity is even worse, washing out the differences between different Hamiltonians and distinguishing only the dimension of the Hilbert space~\cite{Stoica2026TheClockAmbiguityProblemExtendedOrExtinguished}.

Nevertheless, the dynamical stability of the pointer states is a necessary condition, albeit not a sufficient one. Only certain bases can play the role of pointer bases, and they are determined not merely by the dynamical stability or any other structural or relational conditions, but also by their physical meaning. In other words, among the observables, some are special in the sense that they are macroscopically accessible and related more directly to our experience of the physical world. Other observables may satisfy identical relational constraints, but they are not pointer observables. Accounting for the physical meaning of the observables resolves both the ambiguity of decompositions like~\eqref{eq:measurement-problem} and~\eqref{eq:measurement-problem-obs}~\cite{Stoica2022SpaceThePreferredBasisCannotUniquelyEmergeFromTheQuantumStructure,Stoica2026CommentOnOnTheEmergenceOfPreferredStructuresInQuantumTheoryBySoulasFranzmannAndDiBiagio,Stoica2023AreObserversReducibleToStructures}, and the clock ambiguity, in its original or extended version~\cite{Stoica2026TheClockAmbiguityProblemExtendedOrExtinguished}.

Modern readings of Everett's insight consistent with this solution are given, for example, in~\cite{GellMannHartle1990QuantumMechanicsInTheLightOfQuantumCosmology},~\cite{Wallace2012TheEmergentMultiverseQuantumTheoryEverettInterpretation,Saunders2021EverettInterpretationStructure},~\cite{SEP-Vaidman2021MWI}, and~\cite{Stoica2023TheRelationWavefunction3DSpaceMWILocalBeablesProbabilities,Stoica2024ClassicalManyWorldsInterpretation}. But, again, this is not characteristic of Everett's interpretation, decompositions like~\eqref{eq:measurement-problem} and~\eqref{eq:measurement-problem-obs}, induced by macroscopic pointer states, are universal characteristics of all interpretations of quantum mechanics able to account for our macroscopic observations, at least up to the moment when the projection postulate or an alternative resolution is invoked.
At any rate, in the PW proposal Everett's insight is applied to a different problem than the measurement problem.

Let us now return to the PW proposal and the preferred-time problem of clocks introduced in Section~\ref{s:preferred-time-problem}.
Suppose that the evolution of the state of the world $\ket{\psi(t)}_r$ is encoded in the timeless state $\kett{\Psi}=\int_\R\ket{\tau}_c\ket{\psi(\tau)}_r\,\dd \tau$, where $\ket{\tau}_c$ are generalized eigenstates of the operator $\obs{T}_c$.
But, as seen in Section~\ref{s:preferred-time-problem}, under Assumption~\ref{assumption:timelessness}, nothing singles out $\obs{T}_c$ as a time observable over any other choice $\obs{T}_c'=\obs{S}_c\obs{T}_c\obs{S}_c^\dagger$, where $\obs{S}_c$ is a unitary operator on $\hilbert_c$.
Equation~\eqref{eq:PW-world-state-map-new} shows that a generic choice of $\obs{T}_c'$ conditions the state of the world at the time $\vartheta$ to be a generic superposition of states of the world $\ket{\psi(\tau)}_r$ at all times $\tau\in\R$,
\begin{equation}
\label{eq:PW-world-state-map-new-generic}
\ket{\psi'(\vartheta)}_r
=\int_\R c(\tau,\vartheta)\ket{\psi(\tau)}_r\,\dd\tau.
\end{equation}

Therefore, Assumptions~\ref{assumption:timelessness} and~\ref{assumption:clock-conditions} lead to an underdetermination of the instantaneous states, since they do not select states like $\ket{\psi(\tau)}_r$ over generic superpositions of such states.
A generic timeless superposition as in equation~\eqref{eq:PW-world-state-map-new-generic} can be split into definite temporal states only by pointer observables intrinsic to the world's state space $\hilbert_r$.
Just like the superpositions of the form~\eqref{eq:measurement-problem} and~\eqref{eq:measurement-problem-obs} are resolved by macroscopic pointer states into states containing an observed system with a definite spin, in the same way the superpositions of the form~\eqref{eq:PW-world-state-map-new-generic} are resolved into states of definite time.
The systems whose change indicates to us the existence of time are systems within the world.

In practice, time can be tracked by the states of various systems: clocks, calendars, tree rings, planetary systems, the cosmic background radiation, carbon-14, and so on. These timekeepers are not necessarily synchronized to indicate the same time at the same rate. They may be tuned to different time zones or run faster or slower, but for all practical purposes we can synchronize them, or apply the necessary corrections, to determine time with extraordinary precision.
We can, in principle, encode the age of the universe as a high-resolution global time observable $\widehat{\obs{T}}_r$ on $\hilbert_r$.

Suppose we start with a ``wrong'' time observable of the clock system $\obs{T}_c'=\int_{\R}\vartheta\dyad{\vartheta}_c\,\dd \vartheta$.
Then, according to Assumption~\ref{assumption:clock-conditions}, $\obs{T}_c'$ induces the temporal states $\ket{\psi'(\vartheta)}_r$ in $\hilbert_r$.
But if a global time observable $\widehat{\obs{T}}_r$ exists, and if it can distinguish states at different times, it picks out the correct temporal states $\ket{\psi(\tau)}_r\in\hilbert_r$. Then $\widehat{\obs{T}}_r$, not $\obs{T}_c$, is the observable that realizes the decomposition into relative states from equation~\eqref{eq:PW-state-timeless}.
The intrinsic time observable $\widehat{\obs{T}}_r$ induces the correct clock time observable $\obs{T}_c$, and not the other way around.

But maybe $\widehat{\obs{T}}_r$ can distinguish states of the universe at different times only in discrete steps, and its spectrum is, for example, $\Z$, rather than $\R$. Then it induces a discrete time observable for the clock, and equation~\eqref{eq:PW-state-timeless} is replaced with
\begin{equation}
\label{eq:PW-state-timeless-discrete}
\kett{\Psi}=\sum_{\tau\in\Z}\ket{\tau}_c\ket{\psi(\tau)}_r.
\end{equation}

In all cases, the correct conditioning is not $\ket{\tau}_c\mapsto\ket{\psi(\tau)}_r$, but the reversed one,
\begin{equation}
\label{eq:relative-state-r-c-bis}
\ket{\psi(\tau)}_r\mapsto\ket{\tau}_c,
\end{equation}
which invalidates Assumption~\ref{assumption:clock-conditions}.

If the total Hamiltonian is of the form $\obs{H}=\obs{H}_c\otimes\obs{I}_r+\obs{I}_c\otimes\obs{H}_r$, as it is standard in the PW proposal, the world does not know of a separate clock subsystem, and it has no need to know about it.

Moreover, from the point of view of the world, the preferred-time problem manifests as needing to resolve superpositions of the form~\eqref{eq:PW-world-state-map-new-generic}, that is, superpositions of the form
\begin{equation}
\label{eq:preferred-time-superposition}
\int_\R a(\tau)\ket{\psi(\tau)}_r\,\dd\tau.
\end{equation}

Then, why not remove the separate clock, and let $\kett{\Psi}$ be just
\begin{equation}
\label{eq:bootstrap-state-timeless-bis}
\kett{\Psi}=\int_\R\ket{\psi(\tau)}_r\,\dd \tau?
\end{equation}

This still satisfies the Wheeler-DeWitt constraint equation $\obs{H}\kett{\Psi}=0$, because $\int_\R\ket{\psi(\tau)}_r\,\dd \tau$ is invariant under translations with $\tau$~\cite{Stoica2026QuantumTimeWithoutNegativeEnergy}.

Therefore, a separate clock system as in the common reading of the Page-Wootters proposal is redundant and inconsequential for the world system, its time, and its dynamics.

%------------------------------------------------------------%
\section{Reinterpreting the Page-Wootters formalism}
\label{s:PW-reinterpret}

These results do not fully reject the Page-Wootters proposal, only its understanding based on Assumptions~\ref{assumption:timelessness} and~\ref{assumption:clock-conditions}.
We can adapt the formalism of the Page-Wootters proposal in the following way.
First, replace Assumptions~\ref{assumption:timelessness} and~\ref{assumption:clock-conditions} with new principles.

The timeless world $\kett{\Psi}$ is simply the state in equation~\eqref{eq:bootstrap-state-timeless-bis}.
The new principles are the following.
\begin{principle}
\label{principle:timelessness}
The world system is static, it does not presuppose time or dynamics. Time and dynamics emerge solely from the intrinsic time observable $\widehat{\obs{T}}_r$, constructed from intrinsic pointer observables only.
\end{principle}

\begin{principle}
\label{principle:clock-conditions}
The state of the world at time $\tau$ is conditioned on the value of the intrinsic time observable $\widehat{\obs{T}}_r$,
\begin{equation}
\label{eq:intrinsic-time-state-r}
\tau\mapsto\ket{\psi(\tau)}_r.
\end{equation}
\end{principle}

To be more explicit, let $\widehat{\M}=(\obs{M}_1,\obs{M}_2,\ldots)$ be a set of macroscopic commuting observables (as in~\cite{Stoica2026QuantumTimeWithoutNegativeEnergy}). We assume the set $\widehat{\M}$ to be large enough to generate all macroscopic observables, including an intrinsic time observable $\widehat{\obs{T}}_r$.
Let $\{\widehat{\obs{T}}_r\}\cup\widehat{\M}\cup\widehat{\q}$ form a complete set of commuting observables, where $\widehat{\q}=(\widehat{q}_1,\widehat{q}_2,\ldots)$ generate the remaining, nonmacroscopic observables.

In order to reinterpret the PW formalism, let us recall that, in nonrelativistic quantum mechanics, a scalar particle is represented by a wavefunction
\begin{equation}
\label{eq:nrqm-psi}
\ket{\psi}=\int_{\R^3}\braket{x,y,z}{\psi}\ket{x,y,z}\,\dd x\,\dd y\,\dd z,
\end{equation}
which can be expanded in the form
\begin{equation}
\label{eq:nrqm-psi-expanded}
\begin{aligned}
\ket{\psi} 
&=\int_{\R^3}\braket{x,y,z}{\psi}\ket{x}\ket{y}\ket{z}\,\dd x\,\dd y\,\dd z \\
&=\int_{\R}\ket{x}\underbrace{\(\int_{\R^2}\braket{x,y,z}{\psi}\ket{y}\ket{z}\,\dd y\,\dd z\)}_{\ket{\psi(x)}}\,\dd x \\
&=\int_{\R}\ket{x}\ket{\psi(x)}\,\dd x.
\end{aligned}
\end{equation}

This separation of the coordinate $x$ from $y$ and $z$ is based on the isomorphism 
\begin{equation}
\label{eq:square-integrable-iso}
L^2(\R^3)\cong L^2(\R)\otimes L^2(\R^2)
\end{equation}
between Hilbert spaces of square-integrable functions.
The tensor product from equation~\eqref{eq:square-integrable-iso} does not correspond to a decomposition into subsystems, but to a separation of the degree of freedom $x$ from $y$ and $z$.

One may object that the tensor-product decomposition~\eqref{eq:square-integrable-iso} is not quite the same as a decomposition into subsystems, but only a formal decomposition. Two subsystems can be entangled, but can $x$ be entangled with $y$? Let us start with a more intuitive case, can the position of a particle be entangled with its spin? It can, and this is exactly what we are doing when we measure the spin: the magnetic field of the Stern-Gerlach device separates the spin-up and spin-down components of the wavefunction of a silver atom, deflecting them along two different paths that hit different regions of a plate and leave a mark. The mark is the pointer from which the observer infers whether the spin of the silver atom was up or down. So we can separate a degree of freedom from the other ones, and we often do this.

Similar to how we separated $x$ from the other coordinates in equation~\eqref{eq:nrqm-psi-expanded}, we can separate the degrees of freedom given by the possible eigenvalues $(\tau,\M,\q)$ of $\widehat{\obs{T}}_r$, $\widehat{\M}$, and $\widehat{\q}$. Since the configuration space is of the form $\R\times\mathcal{C}$,
\begin{equation}
\label{eq:tau-separate-dof}
\begin{aligned}
\kett{\Psi} 
&=\int_{\R\times\mathcal{C}} \brakett{\tau,\M,\q}{\Psi}\ket{\tau,\M,\q}\,\dd \tau\,\dd \M\,\dd \q \\
&=\int_{\R} \ket{\tau}_c\underbrace{\(\int_{\mathcal{C}}\brakett{\tau,\M,\q}{\Psi}\ket{\M}\ket{\q}\,\dd \M\,\dd \q\)}_{\ket{\psi(\tau)}_r}\,\dd \tau \\
&=\int_{\R} \ket{\tau}_c\ket{\psi(\tau)}_r\,\dd \tau.
\end{aligned}
\end{equation}

Here the subscripts $c$ and $r$ are retained only as tensor-factor labels. They no longer denote a separate clock subsystem and an independently existing rest-of-the-world subsystem.

With this reinterpretation of $\hilbert_c$ and $\hilbert_r$ as corresponding to different degrees of freedom rather than subsystems, everything else in the PW formalism remains unchanged. The state vectors $\ket{\tau}_c$ and $\ket{\psi(\tau)}_r$ no longer represent subsystems, but tensor-product factors corresponding to the degree of freedom $\tau$ and the remaining degrees of freedom $(\M,\q)\in\mathcal{C}$. In particular, the conditional expectation value~\eqref{eq:conditional-expectation-value} is between different degrees of freedom, and not between subsystems. The time observable is intrinsic to the world, and there is no need for a separate system to play the role of a clock.

%------------------------------------------------------------%
\section{Conclusions}
\label{s:conclusions}

We have seen that, if we interpret the PW proposal as time emerging from entanglement and the clock conditioning the temporal states of the world, we run into a new kind of ambiguity that I called \emph{the preferred-time problem of clocks}.
The assumption that time does not exist, but emerges as an observable of the clock system, leaves open an infinite number of possible ways for time to emerge. Each choice of the clock's time operator results in a different decomposition of the timeless state as a superposition encoding the dynamics of the world.
The assumption that the clock's state conditions the temporal states of the world introduces an ambiguous dynamics for the world.
Fortunately, the problem can be resolved in the same way it was found, by realizing that the world's observers know about time from the pointer observables of the world, and not from an inaccessible passive clock. This conclusion is reached following the rationale of relative states introduced by Everett and applied by Page and Wootters to their proposal. However, here I followed an updated version of Everett's idea, which relies on macroscopic pointer observables and applies more generally. 
Fortunately, the Page-Wootters formalism can still be preserved without formal change, provided that we replace the interpretation of the time observable as an observable of a separate clock subsystem with an interpretation based on an intrinsic time observable.

In~\cite{Stoica2026QuantumTimeWithoutNegativeEnergy}, I started from a different problem, noticed by Unruh and Wald~\cite{UnruhWald1989TimeAndTheInterpretationOfCanonicalQuantumGravity}, that if the Hamiltonian is bounded from below, no observable can play the role of time if we require it to be monotonically correlated with the Schr\"odinger time $t$. I bite the bullet and show that, if we drop the monotonicity condition, time observables are possible, resulting in the same intrinsic-time-observable account as the one presented in this article. The route from~\cite{Stoica2026QuantumTimeWithoutNegativeEnergy} is completely different, leading automatically to the Wheeler-DeWitt-type constraint~\eqref{eq:PW-WDW} without assuming gravity.

%------------------------------------------------------------%

\end{document}